\documentclass[aps,prev,twocolumn,preprintnumbers,floatfix,nofootinbib]{revtex4-1}
\usepackage[english]{babel}
\usepackage{bm}
\usepackage{times}
\usepackage{ulem}
\usepackage{listings}
\usepackage{slashed}
\usepackage{color}
\usepackage{appendix}
\usepackage{mathtools}
\usepackage{natbib}
\usepackage{epsfig}
\usepackage{dcolumn}
\usepackage{textcomp}
\usepackage{multirow}
\usepackage{graphicx}
\usepackage{soul}
\usepackage{subfigure}
\usepackage{xcolor}
\usepackage{comment}
\usepackage{amsmath}
\usepackage{hyperref}      % Load hyperref before cleveref

\usepackage{hyperref}
\usepackage[capitalize]{cleveref}

\begin{document}

\title{Updated BBN Bounds on Hadronic Injection in the Early Universe: The Gravitino Problem }

\author{Lucia Angel$^{1,2,3}$}
\email{lucia.correa.717@ufrn.edu.br}
\author{Giorgio Arcadi$^{4,5}$}
\email{giorgio.arcadi@unime.it}
%\email[Corresponding author:]{lucia.correa.717@ufrn.edu.br}
\author{Matheus M. A. Paixão$^{1,2}$}
\email{matheus.mapaixao@gmail.com}
\author{Farinaldo S. Queiroz$^{1,2,3,6}$}
\email{farinaldo.queiroz@ufrn.br}

\affiliation{$^{1}$Departamento de F\'isica, Universidade Federal do Rio Grande do Norte, 59078-970, Natal, RN, Brasil}
\affiliation{$^2$International Institute of Physics, Universidade Federal do Rio Grande do Norte,
Campus Universitario, Lagoa Nova, Natal-RN 59078-970, Brazil}
\affiliation{$^3$Millennium Institute for Subatomic Physics at the High-Energy Frontier (SAPHIR) of ANID, Fern\'andez Concha 700, Santiago, Chile}
\affiliation{$^4$Dipartimento di Scienze Matematiche e Informatiche,
Scienze Fisiche e Scienze della Terra, Universita degli Studi di Messina, Viale Ferdinando Stagno d’Alcontres 31, I-98166 Messina, Italy}
\affiliation{$^5$INFN Sezione di Catania, Via Santa Sofia 64, I-95123 Catania, Italy}
\affiliation{$^6$ Departamento de F\'isica, Facultad de Ciencias, Universidad de La Serena,
Avenida Cisternas 1200, La Serena, Chile}
\begin{abstract}

Late-decaying particles naturally arise in many extensions of the Standard Model, directly impacting key cosmological processes in the early universe, such as Big Bang Nucleosynthesis (BBN). BBN studies often consider only electromagnetic energy injection episodes, but in practice long-lived particles are also amenable to hadronic decays. The latter can greatly alter the predicted abundances of light elements such as $\mathrm{D}/\mathrm{H}$, $Y_p$, ${}^3\mathrm{He}/\mathrm{D}$, and ${}^7\mathrm{Li}/\mathrm{H}$. Incorporating up-to-date measurements, we place constraints on the primordial abundance of long-lived particles as a function of their lifetime. Lastly, we apply our results to the gravitino problem and set bounds on the reheating temperature, which controls the gravitino primordial abundance. 
\end{abstract}

\maketitle

\section{\label{Intro}Introduction}

The standard Big Bang Nucleosynthesis (BBN) \cite{mukhanov2005,kolb2018early,dodelson2020} is a cornerstone of modern cosmology, as its predictions of light nuclei abundances show strong concordance with cosmological observations \cite{Cyburt2016,Fields2020}. Theoretical predictions of the primordial abundances of Deuterium (D), Helium-3 (${}^3\rm{He}$), and Helium-4 (${}^4\rm{He}$), formed in the first minutes after the Big Bang, show excellent consistency with observations of the baryon-to-photon ratio, $\eta = 6.1 \times 10^{-10}$, extracted from CMB measurements \cite{Coc2017,Planck2018V}. We highlight that we adopt this value for the baryon-to-photon ratio throughout this work. Aside from the intriguing contrast between the predicted and observed abundances of Lithium-7 (${}^7\rm{Li}$), commonly referred as the Lithium problem \cite{sbordone2010,Singh2019}, BBN provides a robust cross-check between early-universe physics and observational data.
%A key element in determining the abundances of Deuterium (D), Helium-3 (${}^3\rm{He}$), Helium-4 (${}^4\rm{He}$), and Lithium-7 (${}^7\rm{Li}$), formed in the first minutes after the Big Bang, is the baryon-to-photon ratio $\eta=n_b/n_{\gamma}$, which is tightly constrained by both BBN and CMB data. %The Planck results point to $\eta=6.104\pm0.058\times10^{-10}$, while $\Omega_b h^2={0.02230\pm0.00021}$ for a reduced Hubble parameter $h=0.674$ ***check this h value***.   and Lithium-7 (${}^7\rm{Li}$)

The formation and respective abundances of light element nuclei occur in a well-defined sequence of events during the first few minutes of the Universe, starting when the temperature drops below $~1$ MeV, causing the weak interaction to decouple and setting the neutrons-to-protons abundance to $~1/6$. From this point on, neutrons and protons begin to combine to form the first nuclei. The first significant event is the formation of Deuterium, occurred when the temperature falls below $~0.1$ MeV and the photodisintegration effects become mild \cite{FIELDS2006}. This key process triggers further nuclear reactions and results in the synthesis of other light elements such as ${}^3\mathrm{He}$ and ${}^4\mathrm{He}$, in addition to small amounts of ${}^7\mathrm{Li}$.%${}^6\mathrm{Li}$

The precise values of the abundances, predicted by BBN, rely on the value of $\eta$, the expansion history of the universe (given in terms of the Hubble parameter $H$), and any additional sources of entropy or relativistic degrees of freedom. The change in any of these parameters implicates different abundances, which should be tested with the up-to-date observations of light elements. Therefore, measurements of present-day concentrations provide stringent bounds that can be used to probe non-standard cosmological scenarios, like the ones arising from late-time decays \cite{Kawasaki:2004yh,Kawasaki2018} and additional relativistic degrees of freedom ($N_{\mathrm{eff}}$) \cite{Pospelov2010}. %Such types of models have particular importance since they can handle
These models are particularly important because they have the potential to provide novel insights into dark matter and address important cosmological tensions, like those regarding the Hubble constant \cite{Alcaniz2021,Deivid:2022,Deivid:2023,daCosta2024,deJesus2024} and structure formation \cite{Karananas2018}. These problems should be treated within the allowed limit imposed by BBN (as well as by other cosmological probes), which includes the lifetime of the mother particles and their initial abundances \cite{daCosta2024,deJesus2024}. 

%By using the most updated values of the light elements abundances, we will bring up to date the lifetime and the energy injected from hadronic decays. Additionally, the gravitino case will be considered.

In particular, late-decaying particles can leave imprints in the abundances of light elements by altering the neutron-to-proton ratio, injecting entropy, or introducing high energetic photons and hadrons in the primordial plasma \cite{Kawasaki:2004yh,Kawasaki2005A,Forestell2019}. The presence of high-energy photons generated in the decays can induce photodissociation processes, resulting in the overproduction of lighter nuclei, especially when the photons-energy surpass the nuclei binding energy \cite{Kawasaki2018}. Alternatively, the hadronic showers might provoke scattering processes. For $t \lesssim 100s$, the interconversion of protons and neutrons increases the neutron-to-proton ratio, $n/p$, leading to greater concentrations of $^{4}$He and D in comparison to the standard Big Bang Nucleosynthesis scenario \cite{Kawasaki2005A}. For $t \gtrsim 100s$, on the other hand, the inelastic collisions involving the hadronic jet and the background $^{4}$He implicates in the overproduction of D, $^{3}$He, $^{6}$Li, and $^{7}$Li, which can significantly alter the standard predictions \cite{Kawasaki2005A}. Hence, by comparing the theoretical predictions of the light-element abundances with astrophysical observations, one can set limits on the mass, lifetime, and interaction strength of the new particles responsible for changing the cosmic evolution \cite{Alves:2023jlo}.

%The gravitino is the supersymmetric partner of the graviton \cite{freedman2012},, is a noteworthy example of long-lived particles. 
The gravitino, the supersymmetric partner of the graviton \cite{freedman2012}, is a noteworthy example of such long-lived particles.
Its mass is determined by the specific supersymmetry-breaking model under consideration \cite{Weinberg1982}. For masses in the range $m_{3/2}\sim 10^2-10^3$GeV, the corresponding lifetime exceeds 1s, and its decay can significantly alter the predicted abundances \cite{Kawasaki2005A}. This effect is referred as the gravitino problem in the literature \cite{ELLIS1984,KHLOPOV1984}. As a consequence, observational constraints from primordial element abundances restrict the possible gravitino decay channels, as well its lifetime and its mass.
Notice that a non-neglible abundance of gravitinos at the onset of the BBN is in general expected. Indeed, even if the population of primordial gravitinos can be diluted away during inflation, scattering processes of thermal particles during reheating \cite{Kohri2006,Rychkov2007,Benakli:2017whb} can recreate it. The comoving abundance $Y_{3/2}$ produced in the latter way is an increasing function of the reheating temperature $T_R$ marking the beginning of the standard radiation dominated phase of the history of the Universe. As a matter of fact, in high reheating temperature scenarios, on can have a high enough abundance of gravitinos at the onset of BBN so that the primordial abundances of light nuclei can be dramatically affected by their decays \cite{Kawasaki:2004yh,Kawasaki2005A,Covi:2009bk,Kawasaki2018}. BBN constraint can be hence used to indirectly constrain the reheating temperature after inflation.
As can be easily guessed, the gravitino problem occurs in scenario in which the gravitino is not the the lightest supersymmetric particle. According the supersymmetic model gravitino can be the cosmologically stable Lightest Supersymmetric particle (LSP), and possibly a DM candidate. Most of the findings of this paper would apply also in such a setup. Indeed the decays of the Next-to-Lightest supersymmetric particle into gravitino and SM products would occur a timescales comparable to the onset of BBN, because of the Planck scale suppression of the interactions between NLSP and LSP, and have an analogous impact on light nuclei abundances as the decay of the gravitino (see e.g. \cite{Covi:2009bk}.) 

%In inflationary contexts, even with the dilution effect of the primordial gravitinos caused by the inflation, the same is produced by scattering processes of thermal particles during the reheating \cite{Kohri2006}. Consequently, it is possible to relate the gravitino abundance $Y_{3/2}$ with the reheating temperature $T_R$. As a matter of fact, in models with high-reheating temperatures, $Y_{3/2}$ could become large enough and substantially alter the concentrations of light elements through gravitino decays \cite{Kawasaki:2004yh,Kawasaki2005A,Covi:2009bk,Kawasaki2018}. This fact imposes stringent limits to the reheating temperature after inflation. 

In this paper, we will not give a detailed explanation of the chain reactions involved in the formation of light elements and how the late-time decays changes the standard BBN. For technical details, the reader is encouraged to consult \cite{Kawasaki:1994sc}. The focus is to present the latest BBN constraints regarding the hadronic channels involved in late decays. Lastly, we will apply our findings to the gravitino model, and set bounds on the reheating temperature which controls the gravitino primordial abundance. 

%In particular, late-decaying particles can modify the standard BBN predictions by altering the neutron-to-proton ratio, injecting entropy, or introducing energetic hadrons and photons capable of dissociating nuclei after their formation. These effects can leave observable imprints on the primordial abundances of light elements, providing a unique window into physics beyond the standard model

%Consequently, the BBN put stringent bounds on the observed abundance of light-elements, serving as an important probe to non-standard cosmological scenarios. 
%One recurrent example is the energy injection at early times due to the decay of long-lived particles with lifetimes $\tau>1\rm{s}$. If such particles are massive enough, they can induce hadronic and electromagnetic interactions with the primordial plasma, leading to a cascade of events that eventually alters the original abundances, with photodissociation and hadrodissociation processes playing an important role. To a detailed explanation about the BBN and the effect of late-decay particles see Refs [??,??].

In brief, the work is organized as follows: In \autoref{Abundance}, a short summary of the primordial abundances of light elements is presented. In \autoref{Discussion}, the essential aspects required for calculating the hadronic decay are discussed and the constraints on the abundances are established. The \autoref{Gravitinosec} addresses the gravitino problem. Finally, in \autoref{Conclusions}, the conclusions are outlined.

\noindent

\section{Abundance of the light elements \label{Abundance}}

The distribution of light elements is determined by solving the first-order differential Boltzmann equations \cite{Tytler:2000qf,Pospelov:2010hj},
\begin{eqnarray} \label{start1} \frac{dY_i}{dt} = -H(T)T \frac{d Y_i}{dT} = \sum (\Gamma_{ij}Y_j + \Gamma_{ikl} Y_k Y_l + \dots), \end{eqnarray} where $Y_i = n_i / n_b$ represents the ratio of the number density of species $i$ to the baryon number density. The interaction and decay rates are represented by $\Gamma_{ij}$ and $\Gamma_{ikl}$, respectively. The Hubble expansion rate, $H(T)$, is given by,

\begin{equation} \label{Hubbleeq} H(T) = T^2 \left(\frac{8\pi^3 g_* G_N}{90}\right)^{1/2}, \end{equation} where $g_* = g_{\text{bosons}} + \frac{7}{8}g_{\text{fermions}}$ corresponds to the relativistic degrees of freedom at the time, $G_N$ is the gravitational constant, and $T$ is the plasma temperature.

%Eq.~\eqref{start1} requires consideration of the differing neutrino and photon temperatures after electron-positron annihilation, where $T_\nu \simeq (4/11)^{1/3} T_\gamma$. 

Several processes play an important role in the synthesis of light elements. We briefly describe them below: \\

\paragraph{\textbf{Helium-4}} The conversion of neutrons to protons is governed by weak interactions, with rates scaling as $\Gamma_{n-p} \sim G_F^2 T^5$, where $G_F$ is the Fermi constant. As the universe cools, with $H(T) \sim \sqrt{g_* G_N}T^2$, the interaction rate decreases more rapidly than the Hubble rate. This leads to freeze-out near $T \sim 1$~MeV, at which point the neutron-to-proton ratio is given by $n/p \sim \exp^{(m_n - m_p)/1 \text{ MeV}} \sim 1/6$, which is further reduced to $1/7$ due to $\beta$ decay. All neutrons are incorporated into Helium-4, because the abundance of background photons prevents deuterium formation via photodissociation. Consequently, the Helium-4 mass fraction, $Y_p$, is mostly determined by the neutron-to-proton ratio \cite{Pospelov:2010hj},

\begin{equation} Y_p \simeq \frac{2n/p}{1 + n/p} = 0.248. \end{equation}

Thus, the Helium-4 fraction is influenced by the freeze-out neutron-to-proton ratio and the onset of efficient deuterium production.\\

\begin{figure*}[ht!]
\centering
    {
    \includegraphics[width =1.8 \columnwidth]{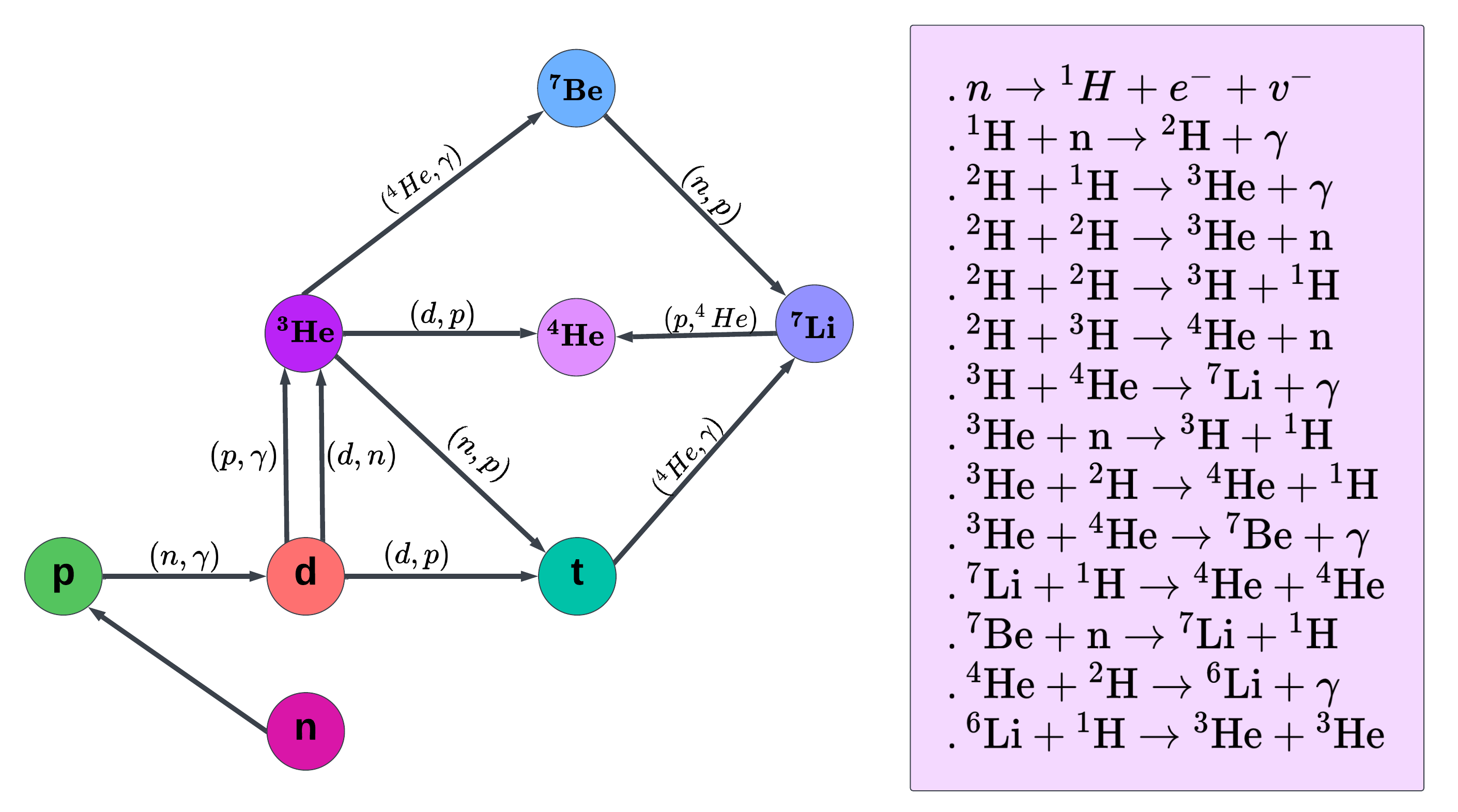}
    \caption{Diagram representing the chain reactions involved of production of light elements during BBN.} 
    \label{fig:chains}
    }
\end{figure*}

\paragraph{\textbf{Deuterium:}} When the temperature drops below $70$~keV, the suppression of photons due to the Boltzmann factor becomes significant, allowing for substantial deuterium (D) production and triggering subsequent nuclear reactions (see \autoref{fig:chains}). Deuterium and $^3$He are synthesized simultaneously, with most deuterium reacting with protons and neutrons to form helium and trace amounts of tritium. A minor contribution to the formation of Lithium-7 is also provided by deuterium.\\

\paragraph{\textbf{Lithium-7:}} The production rates for elements with $A=6,7$ are generally lower than the Hubble expansion rate, resulting in low abundances. Approximately 90\% of primordial $^7$Li is produced from $^7$Be as indicated in \autoref{fig:chains}. We acknowledge that the primordial abundance of Lithium-7 is debated, but we will include it in our analysis. Thus, one should bear in mind, the constraints derived based on Lithium-7 should be taken with a grain of salt.\\

\paragraph{\textbf{Lithium-6:}} Due to significant uncertainties, the primordial abundance of $^6$Li is not commonly used to constrain new physics. Observations in halo stars \cite{Asplund:2005yt} indicate a $^6$Li/$^7$Li ratio of approximately $0.05$. While $^6$Li could be used to impose constraints on new physics, challenges in determining its primordial abundance have led to its exclusion from this analysis.

That said, we proceed to compute the influence of hadronic decays in the abundance of light elements. The updated abundances of the light elements were taken from \cite{Alves:2023jlo}.

\section{\label{Discussion}Decay Modes}

We are interested in the decaying scenario where a massive particle $\chi$ with mass $m_X$ and decay rate $\Gamma_\chi$ possesses a non-zero number density in the early universe. The production mechanisms for $X$ depend on its intrinsic properties; however, this study does not focus on the specific production processes. Consequently, the results presented are model-independent and broadly applicable, as shown in Section \ref{Gravitinosec}. Our constraints rely on the lifetime of the decaying particle, its mass, and relic density prior to decay, which is parametrized in terms of the Yield, $Y_X$, defined as,

\begin{equation}
    Y_\chi \equiv \frac{n_\chi}{s},
\end{equation}
where $n_\chi$ denotes the number density of $\chi$, and $s$ represents the total entropy density of the universe. This variable is defined at times $t \ll \tau_\chi$, where $\tau_\chi=\Gamma_\chi^{-1}$. Assuming negligible entropy production, $Y_\chi$ remains constant during this period. Notice that the mass of the decaying particle and final state sets the amount of energy deposited into the medium. The effect that this energy deposition causes in the medium highly depends on when this deposition occurred.

The energy injection episodes caused by long-lived particles are often categorized as electromagnetic or hadronic \cite{Tytler:2000qf}. However, even electromagnetic decays may induce hadronic modes \cite{Kawasaki:2004yh, ParticleDataGroup:2022pth}. For this reason, a hadronic branching ratio is introduced with $B_h = \Gamma_{\chi\rightarrow \text{hadrons}}/\Gamma_\chi$, where $\Gamma_\chi$ is the total decay rate of the particle $\chi$.

If $\chi$ decays directly into colored particles, $B_h$ approaches unity. Even if $\chi$ decays predominantly into photons (or other non-hadronic particles), $B_h$ may still be non-zero due to possible quark-antiquark pair production from virtual photon lines, inducing non-trivial and sizable changes in the light element abundances \cite{Kawasaki:2004yh}. We properly account for this hadronic decay using the energy of the primary parton jet, $E_{\text{jet}}$, which for $B_h=1$, $E_{\text{jet}} = m_\chi / 2$ following the method described \cite{Kawasaki:2004qu}.

Anyway, the literature has devoted more attention to electromagnetic injections at early times, but in this work, we will focus on hadronic episodes motivated by the gravitino problem and other new physics studies \cite{Cerdeno:2005eu,Jedamzik:2007cp,Bailly:2010hh,Pradler:2007is,Hooper:2011aj,Kelso:2013paa,Kelso:2013nwa,Allahverdi:2014bva,Hufnagel:2017dgo,Alves:2023jlo,Sobotka:2023bzr}. In the following sections, we detail the reasoning behind these energy injections.

\subsection{Electromagnetic Injections}

In electromagnetic injections, the decaying particle produces photons and electrons. The most important process that governs the destruction rate of the light elements is photon-photon process where the high energy photons originated from the decaying particle scatter off the background photons producing $e^+e^-$ pairs. This process quickly thermalizes the high energy photons yielding no meaningful change in abundance of light elements. However, low energy photons produced in the cascade that do not have sufficient energy to induce $e^+e^-$ pair production, i.e. with $E_\gamma <  m_e/22T$, can still destroy light elements such as $^4$He, and other elements greatly altering the BBN predictions \cite{Kawasaki:2004yh, ParticleDataGroup:2022pth}. Photodissociation processes become effective at temperatures below $T \sim 0.01$ MeV for D and $T \sim 0.001$ MeV for $^{4}\text{He}$.  

For $\tau < 10^6$s, constraints arise from the destruction of D\cite{Kawasaki:2004yh}, whereas for $\tau > 10^6$s the overproduction of D and $^{3}\text{He}$ due to $^{4}\text{He}$ destruction dominates.  

\subsection*{Hadronic Injection}

When quarks or gluons are emitted from the decaying particle, they fragment into hadrons, forming hadronic jets. High-energy mesons and
nucleons injected into the plasma can alter nucleosynthesis
predictions. When quarks or gluons are emitted, they fragment into hadrons, forming hadronic jets. High-energy mesons and nucleons injected into the plasma can alter nucleosynthesis predictions \cite{Kawasaki:2004yh}.  

\begin{itemize}
    \item At early times ($t \lesssim 100$ sec), mesons and nucleons lose energy via electromagnetic interactions, reaching kinetic equilibrium without directly destroying light elements. However, they can interconvert background protons and neutrons, altering the $n/p$ ratio and increasing the production of $^{4}\text{He}$ and D.  
    \item At later times ($t \gtrsim 100$ sec), mesons decay before interacting with nucleons. High-energy nucleons can, however, scatter off background nucleons and $^{4}\text{He}$, producing secondary hadrons that evolve into hadronic showers. These showers destroy $^{4}\text{He}$ and produce D, T, $^{3}\text{He}$, $^{6}\text{Li}$, and $^{7}\text{Li}$.  
\end{itemize}

Beyond these standard processes, recent studies \cite{Kawasaki_2018, Akita_2025, Akita:2024nam} have highlighted important refinements in hadron-nucleon interactions, particularly regarding n-p interconversion effects and the role of long-lived metastable particles like pions and kaons. While these effects are most significant for $\tau < 1$s (at the boundary of our excluded region), our analysis focuses on $\tau > 1$s where hadronic showers dominate light element production. For $\tau \sim 1$s, proton-neutron interconversion drives $^4$He overproduction, while at shorter timescales ($\tau < 0.1$s) the non-thermal processes become inefficient, requiring substantially larger energy injections. Although some studies extend constraints to $\tau \sim 0.02$s \cite{Boyarsky_2021,Fradette:2018hhl} through neutrino dynamics or analytic methods, such early decays involve additional complexities beyond our present scope.
\begin{figure*}[ht!]
\centering
    \subfigure[]{
    \includegraphics[width =1 \columnwidth]{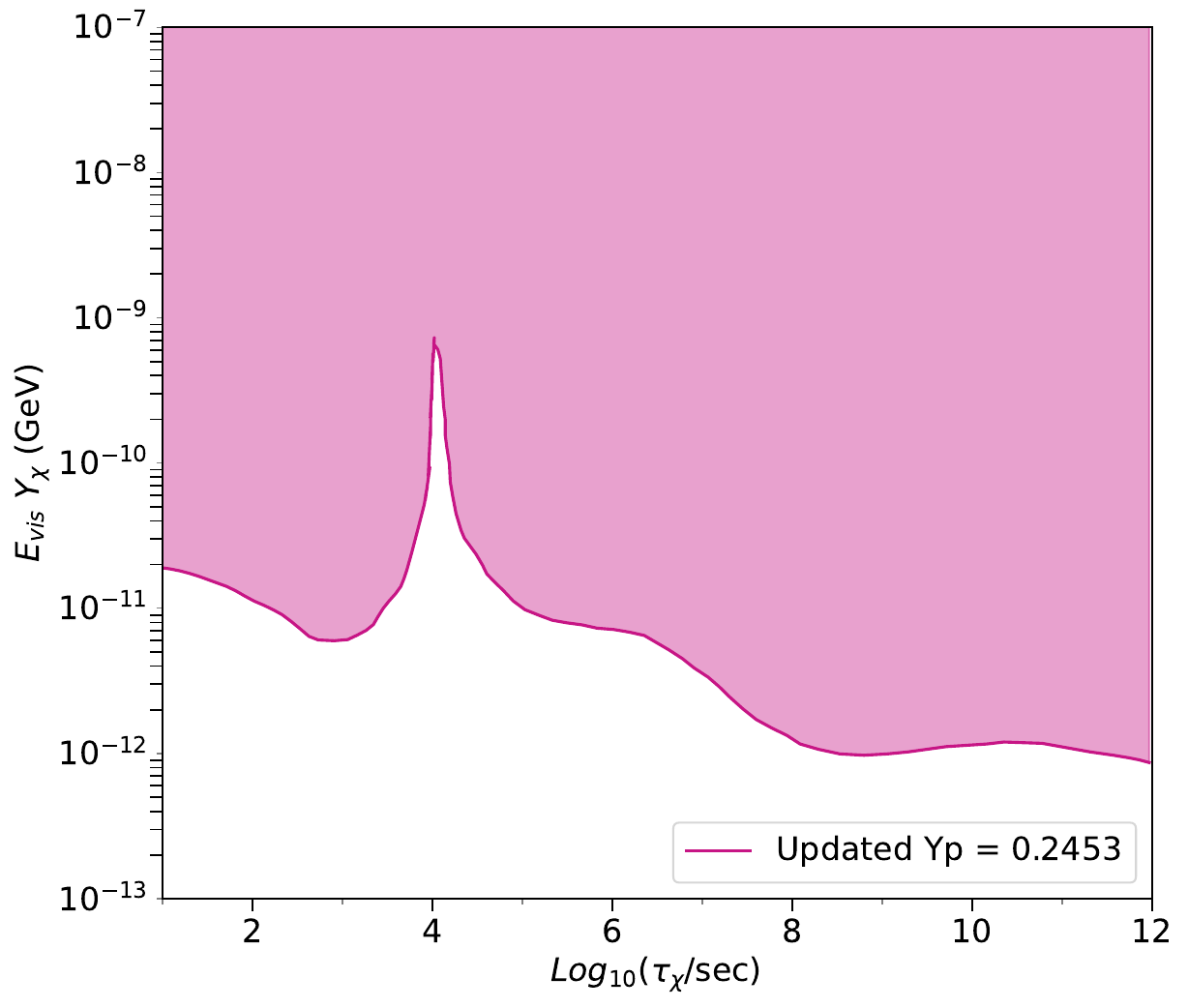}
    \label{fig:a}
    }
    \subfigure[]{
    \includegraphics[width =1 \columnwidth]{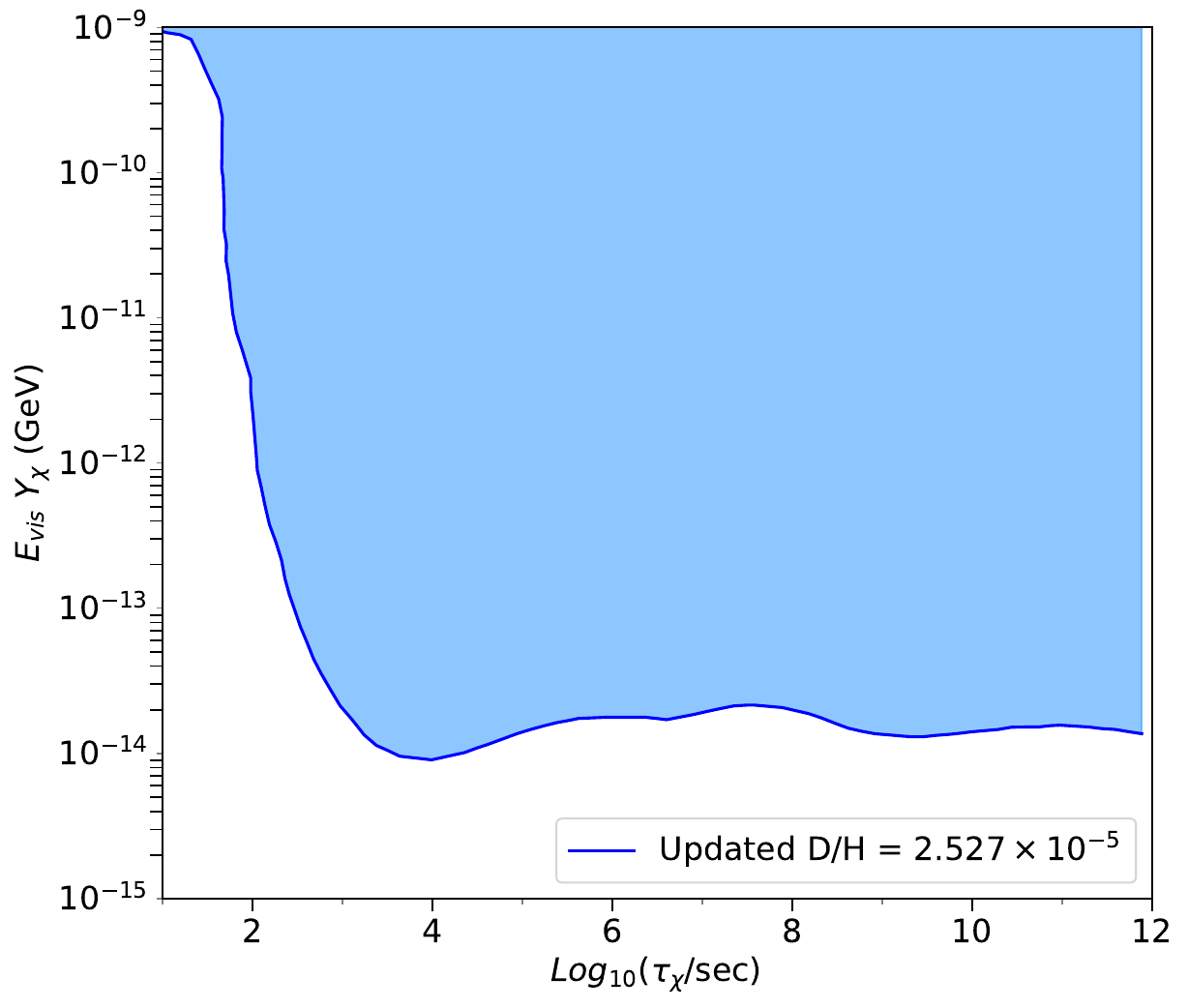}
    \label{fig:b}
    }
         \caption{Exclusion regions based on the $Y_p$ (Figure \autoref{fig:a}) and D/H (Figure \autoref{fig:b}) primordial abundances. The exclusion counters are shown in the $E_{\mathrm{vis}} Y_\chi \times \tau_\chi$ plane, for $m_\chi = 1$ TeV.} 
    \label{fig:Upper bounds}
\end{figure*}

\begin{figure}[ht!]
    \centering
    \includegraphics[width=1\linewidth]{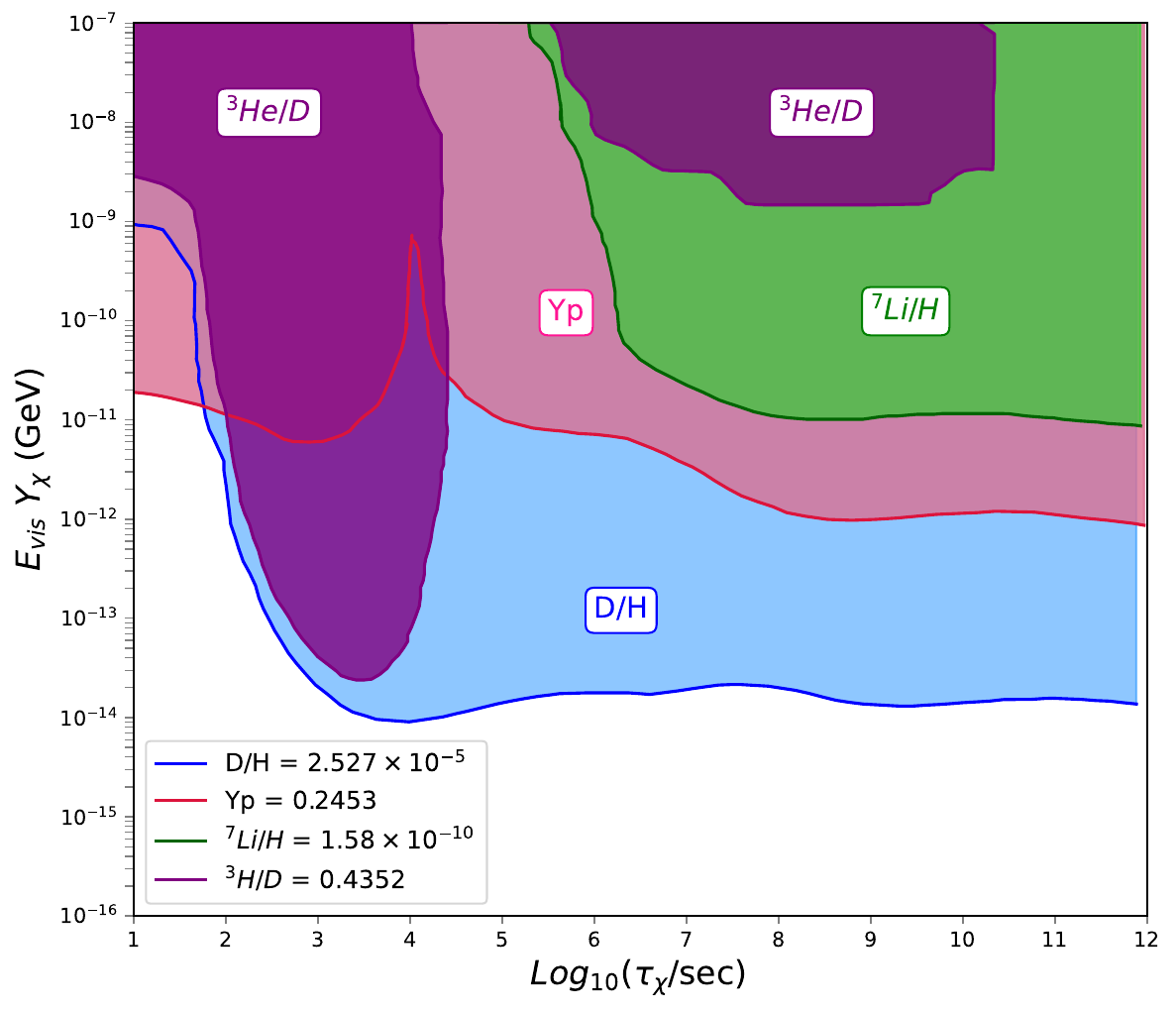}
    \caption{BBN bounds on hadronic injection based in the $E_{\mathrm{vis}} Y_\chi$ vs $\tau_\chi$ plane. These limits rely on the primordial abundances of $Y_p$ (red), D/H (blue), $^3\mathrm{He}/\mathrm{D}$ (purple), and $^7\mathrm{Li}/\mathrm{H}$ (green) as indicated in the figure. We assumed $m_\chi = 1$ TeV and $B_h = 1$. Thus, $E_{\mathrm{vis}} = m_\chi$.}
    \label{fig:bounds}
\end{figure}

Hadronic decays impose stringent constraints due to non-thermal light element production, even at $t \lesssim 10^6$ seconds when radiative decay becomes less significant. Notably, most of the energy from primary hadrons is transferred to electrons and photons via electromagnetic interactions, resulting in similar effects to radiative decay \cite{Kawasaki:2004yh}. 

This work provides an updated analysis of constraints on hadronic decays of massive particles and applies these results to the gravitino. The analysis incorporates the most recent observational data on light element abundances \cite{ParticleDataGroup:2022pth}, offering refined limits that improve upon previous studies\cite{Kawasaki:2004yh}.  

%%%%%%%%%%%%%%%%%%%%%%%%%%%%%%%%%%%%%%%%%%%%%%%%%%%%%%%%%%%%%%%%%%%%%%%%%%%
The hadronization processes are first considered. Since the mass of $\chi$ is assumed to be larger than the QCD scale, $\chi$ primarily decays into quarks and gluons during the hadronic decay process. For the cosmic temperature under consideration, the time scale of hadronization is significantly shorter than that of cosmic expansion. Consequently, partons emitted from decay of $\chi$ are instantaneously hadronized and fragmented into mesons and nuclei, such as $\pi^{\pm}$, $\pi^0$, $K^{\pm}$, $K^0_{L,S}$, $n$, $p$, and $\Lambda^0$, among others.

In the analysis of cascade processes, these energetic nuclei and mesons (particularly $p$ and $n$) are identified as the primary particles responsible for initiating the hadronic shower. For the purposes of this study, the uncertainties associated with the hadronization processes are considered negligible.\\
To simulate these hadronization products, we employ JETSET 7.4 \cite{Kawasaki:2004qu}, with updated cross sections for BBN interactions taken from \cite{Serpico_2004,Cyburt_2008}. Although modern generators such as Pythia8 offer alternative implementations \cite{Kawasaki_2018}, our approach using JETSET 7.4 with updated observational constraints provides robust results consistent with current abundance measurements \cite{Alves:2023jlo}. The predicted abundances were computed as functions of $\eta$ and verified against observations for $\eta = 6.1 \times 10^{-10}$, following a methodology similar to \cite{Kawasaki_2018} but with updated primordial abundances from \cite{Alves:2023jlo}.

When analyzing the effects of the hadronic decay of $\chi$, the spectra of the primary hadrons and the corresponding total energy of the jets are determined by the mass of $\chi$ and its decay modes. For instance, for $m_\chi \sim \mathcal{O}(100)\ {\rm GeV} - \mathcal{O}(100)\ {\rm TeV}$, the typical energy of the produced hadrons is estimated to be approximately $\mathcal{O}(1-100)\ {\rm GeV}$ \cite{Kawasaki:2004yh}.

To study the behavior of light element abundances, these were estimated using the central values of cross sections and model parameters. The contours of $\mathrm{D}/\mathrm{H}$, ${}^3\mathrm{He}/\mathrm{D}$, $Y_{\rm p}$, and ${}^7\mathrm{Li}/\mathrm{H}$ in the ($\tau_{\chi}$, $m_{\chi}Y_{\chi}$) plane are shown in \autoref{fig:bounds}. In the calculations, the effects of photodissociation and hadrodissociation are included. We take $m_\chi$ = 1 TeV and $B_h = 1$, considering the case where $\chi$ decays into two hadronic jets with energy $2E_{\rm jet} = m_\chi$.

As shown in \( Y_p \) (see Fig. \autoref{fig:a}), for \( \tau_{\chi} < 10^0 \, \mathrm{s} \), unstable particles impact in the proton-neutron equilibrium by liberating neutrons, leading to an increase in \( Y_p \) abundance. During \( \tau_\chi \sim 10^1 - 10^2 \, \mathrm{s} \), energy injection plays a critical role, as neutrons are actively fusing into \( Y_p \). In this timeframe, hadronic decay products contribute moderately to \( Y_p \) formation, compared to their impact at earlier stages.

For \( \tau_\chi \gtrsim 10^{3} - 10^{4} \, \mathrm{s} \), the abundance of \( \mathrm{^4He} \) begins to decrease as the primordial abundance of \( \chi \) increases. This behavior is attributed to the efficiency of hadrodissociation and photodissociation processes, which lead to the destruction of \( \mathrm{^4He} \) nuclei. These destruction processes are coupled with the production of lighter elements, such as \( \mathrm{D} \), \( \mathrm{^3He} \), and \( \mathrm{^6Li} \), whose abundances initially rise with increasing \( Y_\chi \).

At longer decay times (\( \tau \geq 10^4 \, \mathrm{s} \)), the formation of \( Y_p \) remains unaffected, as BBN has concluded and temperatures are no longer conducive to nuclear reactions. High visible energy contributions (\( E_{vis} Y_{\chi} \)) can still impact \( Y_p \) through decay products that promote its destruction, particularly in earlier epochs where \( Y_p \) is more sensitive to such effects. 

Now, in the Fig. \autoref{fig:b}, we can see that for earlier times, the deuterium has not been formed yet, so the impact of $\chi$ in this case is limited. During $10^0$s $\leq \tau_\chi \leq 10^2$s we can see an evident diminution of $\mathrm{D}/\mathrm{H}$. Considering high energies in these frames of time shows us  how the $\mathrm{D}/\mathrm{H}$ is sensible to the photodissociation produced by this excess of energy. After this range, the impact of hadronic decays are negligible.  

Analyzing the behavior of ${}^3\mathrm{He}/\mathrm{D}$ contour showed in \autoref{fig:bounds} in purple, we can see that during $\tau \leq 10^2$s, the formation is simultaneous with the formation of the Deuterium, so depending on the model, the hadronic injection of energy can stimulate either the production or destruction of those isotopes, affecting their differential ratio. For $\tau \geq 10^4$s, the hadronic injection occurs too late to cause a big impact in the ${}^3\mathrm{He}/\mathrm{D}$ ratio. 

As shown in green in \autoref{fig:bounds}, for \( ^7\mathrm{Li}/\mathrm{H} \), low values of \( E_{\mathrm{vis}} Y_\chi \) result in an abundance close to the standard prediction. However, at higher energies, the energetic injection can destroy \( ^7\mathrm{Be} \) and consequently disrupt its production chain, significantly reducing its abundance for \( 10^0 \leq \tau_\chi \leq 10^4 \, \mathrm{s} \). This mechanism could potentially justify the gap between the theoretical and observed prediction of $^7\mathrm{Li}$ abundance \cite{Alcaniz:2019kah,Goudelis:2015wpa}.

\section{\label{Gravitinosec} Application to Gravitino Problem}

\begin{comment}
\begin{figure}[h!t]in the figure (a)
    \centering
    \includegraphics[width=\columnwidth]{gg.pdf}
    \caption{Upper limits of the reheating temperature, $T_R$, as a function of the gravitino mass, $m_{3/2}$, for the gluon-gluino channel. The curves in blue, red, black, and green were constructed based on the measurements of D/H, $Y_{p}$, ${}^{3}$He/D, and ${}^{7}$Li/H, respectively. The dashed lines represent the constraints obtained from the most recent observations. The main differences are observed in the curves of ${}^{3}$He and ${}^{7}$Li.}
    \label{fig:gg1}
    \includegraphics[width=\columnwidth]{ff.pdf}
    \caption{Constraints on the reheating temperature as a function of the gravitino mass, analyzed for the case where photon-photino is the dominant decay channel. As before, the differences between the newest and oldest observations of D/H and $Y_{p}$ are mild. Nevertheless, the updated observations of the excess of ${}^3$He abundance (dashed-gray line) impose tighter limits on $T_R$ for masses between 100 and 2000 GeV.}
    \label{fig:ff1}
\end{figure}
\end{comment}

In this section, we discuss the gravitino model as a concrete example of late time decay. In supergravity, the gravitino acquires mass as a result of supersymmetry breaking. The gravitino interacts with SM particles inversely proportional to the Planck scale. For this reason, it is naturally a long-lived particle \cite{Weinberg1982}. One of the main decays are into hadrons, and we studied before they have profound implications for BBN. 

Gravitinos may be produced in the early universe nonthermally from the inflaton decays \cite{KAWASAKI2006,Ellis2016}, or much later, after BBN, from the decay of massive unstable particles \cite{Rychkov2007}, and lastly thermally through a freeze-in production mechanism dictated by the reheating temperature $T_{R}$ \cite{Eberl:2020fml}. Despite the possibility of gravitino production through different processes in the early universe, a key quantity that directly impacts the BBN is its efficiency, which is significantly affected by the temperature during reheating \cite{Kawasaki1995}. 

The precise relation between the primordial abundance of the gravitinos $Y_{3/2}$ and the reheating temperature was obtained in \cite{Kawasaki2005A} by means of the Boltzmann equation,
\begin{align}
    \frac{dn_{3/2}}{dt}+3Hn_{3/2}=\langle \sigma_{\text{tot}} v_{\text{rel}} \rangle n^2_{\text{rad}}. \label{Boltzmann}
\end{align}
Here, $n_{3/2}$ is the gravitino number density and $n_{\text{rad}}=\frac{\zeta(3)}{\pi^2}T^3$ is the number density related to the radiation content. The Hubble parameter $H$, given in terms of the total energy density, accounts for the dilution effect of gravitinos due the expansion, while $\langle \sigma_{\text{tot}} v_{\text{rel}} \rangle$, the thermally averaged product of the total cross-section $\sigma_{\text{tot}}$ and the relative velocity of the particles $v_{\text{rel}}$, quantifies the efficiency of processes involving gravitinos, which are mainly due to two-body processes associated to radiation particles. %The yield variable $Y_{3/2}$ is related with $n_{3/2}$ through the definition \begin{align}
% Y_{3/2}\equiv\frac{n_{3/2}}{s},
%\end{align}
%with the entropy density $s=\frac{2\pi^2}{45}g_{*S}T^3$ expressed in terms of the effective number of relativistic degrees of freedom $g_{*S}$ and the temperature $T$. 
For the reheating temperature, the following definition was used \cite{Kawasaki2005A},
\begin{align}
    T_R\equiv \left(\frac{10}{g_{*S}\pi^2}M_{Pl}^2\Gamma_{\phi}^2\right)^{\frac{1}{4}},
\end{align}
where $M_{Pl}$ is the Planck mass, $\Gamma_{\phi}$ is the decay rate associated with the inflaton field, and $g_{*S}$ is the effective number of relativistic degrees of freedom.

By tracking the evolution of $n_{3/2}$, given in terms of the Boltzmann equation \eqref{Boltzmann}, concurrently with the evolutions of inflationary and radiation energy densities from the moment when $H\gg \Gamma_{\phi}$ up to $H\ll \Gamma_{\phi}$, the gravitino abundance can be approximated by the following fitting formula \cite{Kawasaki2005A}, 
\begin{align}
    &Y_{3/2}\simeq1.9\times10^{-12}\left(\frac{T_R}{10^{10}\mathrm{GeV}}\right)\times\label{Y32}  \\
    &\left[1+0.045\ln{\left(\frac{T_R}{10^{10}\mathrm{GeV}}\right)}\right]\left[1-0.28\ln{\left(\frac{T_R}{10^{10}\mathrm{GeV}}\right)}\right].\nonumber
\end{align}

In this calculation, the contribution of all supersymmetric particles were considered \cite{Kawasaki2005A} assuming the gauginos to be lighter than the gravitino. When the gaugino substantially heavier than the gravitino, the gravitino production is dominated by $SU(3)_C$ processes and the result should be rescaled by a factor of $3m_{3/2}^2/m^2_{\Bar{g}_3}$, where $m_{\Bar{g}_3}$ is the corresponding gaugino mass. From Eq.\eqref{Y32}, we conclude that the gravitino density grows with the reheating temperature. As BBN is sensitive to the primordial abundance of the decaying particle at a given lifetime, we can use BBN to place an upper bound on the reheating temperature.

Previous studies have explored different channels of the gravitino decay \cite{Kawasaki2018}. In this paper, we focus on the gluon-gluino ($g\Bar{g}$) and photon-photino ($\gamma\Bar{\gamma}$) channels (where the photino is assumed to be the lightest neutralino). In both cases, the lifetime of the gravitino scales with $m_{3/2}^{-3}$ with \cite{Kawasaki2005A},
\begin{align}
    \tau_{3/2} &= 6 \times 10^7 \, \mathrm{sec} \times \left( \frac{m_{3/2}}{100 \, \mathrm{GeV}} \right)^{-3}, \hspace{2mm} \psi_{\mu} \rightarrow g + \tilde{g},\\
    \tau_{3/2} &= 4 \times 10^8 \, \mathrm{sec} \times \left( \frac{m_{3/2}}{100 \, \mathrm{GeV}} \right)^{-3}, \hspace{2mm} \psi_{\mu} \rightarrow \gamma + \tilde{\gamma}.
\end{align}

For the final state $g + \tilde{g}$ the hadronic branching ratio is large, $B_h=1$. However, if this decay is kinematically suppressed, the main contribution comes from the $\psi_{\mu} \rightarrow \gamma + \tilde{\gamma}$ production. In this case, because virtual photons can generate gluon-gluino pairs, the hadronic branching ratio still has a contribution of the order of $B_h=10^{-3}$. Despite its low value, such processes can significantly contribute to hadronic cascades and should be accounted for \cite{Kawasaki2005A}.

\begin{figure}%[ht!]
\centering
    \includegraphics[width =1 \columnwidth]{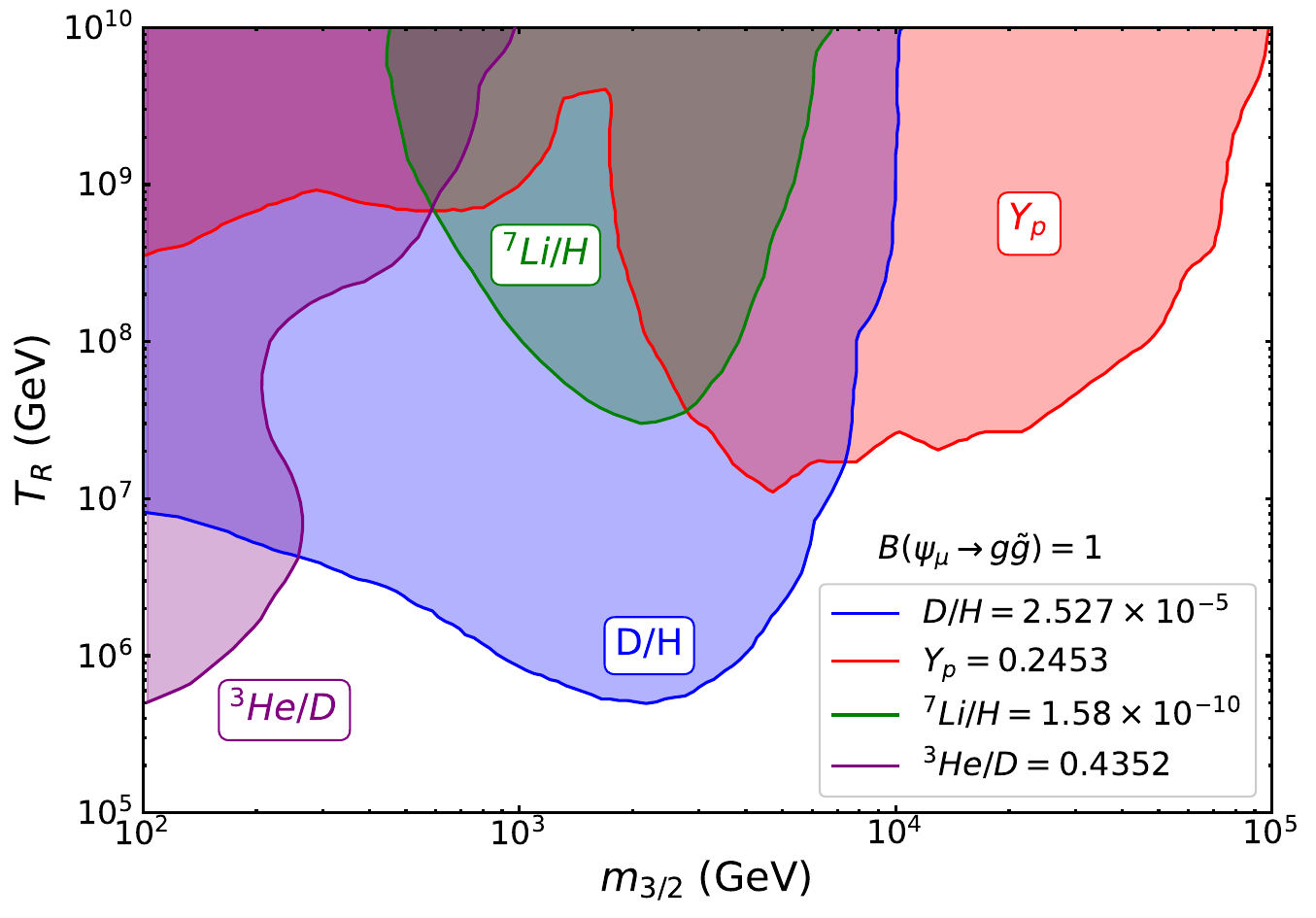}
     \caption{Upper limits of the reheating temperature, $T_R$, as a function of the gravitino mass, $m_{3/2}$, for the gluon-gluino channel (a) and for the photon-photino channel (b). The curves in blue, red, purple, and green were constructed based on the measurements of \rm{D/H}, $\rm{Y_{p}}$, $\rm{^{3}He/D}$, and $\rm{^{7}Li/H}$, respectively. The filled areas represent the constraints obtained from the most recent observations of light element abundances (solid lines). The main differences are observed in the curves of $\rm{^{3}He}$ and $\rm{^{7}Li}$. In particular, figure (b) shows that the $\rm{^3He}$ updated observations impose tighter limits on $T_R$ for masses between 100 and 2000 GeV.} 
    \label{fig:gluon-gluino}
\end{figure}

\begin{figure}%[ht!]
\centering
\includegraphics[width =1 \columnwidth]{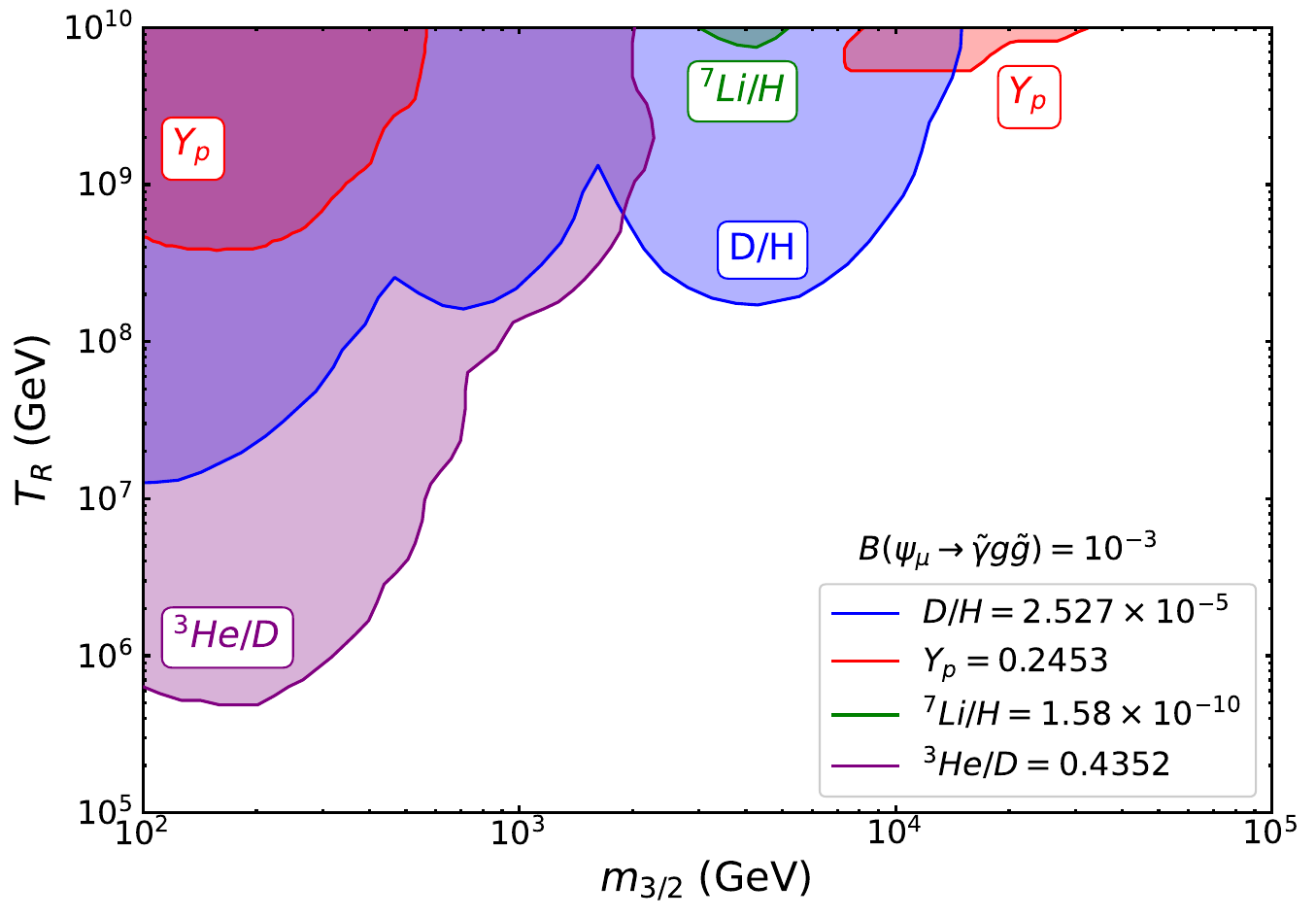}
         \caption{Curves for the photon-photino channel. As in the gluon-gluino case, the differences between the previous and updated observations are more noticeable for ${}^3$He/D and ${}^7$Li/H.}
    \label{fig:foton-fotino}
\end{figure}

In \autoref{fig:gluon-gluino} and \autoref{fig:foton-fotino}, the updated limits on the reheating temperature as function of the gravitino mass are exhibited for gluon-gluino and photon-photino channels, respectively. The hashed areas represent the exclusion regions. In total, four different limits were constructed for each channel, representing the observations of \rm{D/H}, $\rm{Y_p}$, $\rm{^3He}$/\rm{D}, and $\rm{^7Li/H}$. Relevant differences for the constraints arising from \rm{D/H}, $\rm{^3He/D}$, and $\rm{^7Li/H}$ are identified, while for $Y_p$, no considerable deviations were observed. For the gluon-gluino case, the primary constraint on the reheating temperature for masses below the weak scale stems from the excessive production of $\rm{^3He}$, with a reheating temperature $T_R \lesssim  5\times 10^6$ GeV.  For masses below 7 TeV (but greater than the weak scale), non-thermal production of D imposes tighter constraints, with the uppermost limits on the reheating temperature varying from 5$\times10^5$ GeV to 2$\times10^7$ GeV. After that, the interconversion effect of protons and neutrons becomes significant, enhancing the neutron's abundance and boosting the production of $\rm{^4He}$. Consequently, $T_R$ can be relaxed until $10^{10}$ GeV for masses bellow $100$ TeV. 

\noindent A similar situation arises for the photon-photino channel. The main difference lies in the greater significance of the $\rm{^3He}$ overproduction across a greater mass range, with the restrictions imposed in the reheating temperature overcoming the limits of D excessive production up to $m_{3/2}\lesssim$ 2 TeV. Within this interval, the updated limits on $T_R$ range from 5$\times10^5$ GeV to 7$\times10^8$ GeV. Beyond this mass value, the excessive production dominance shifts to D and, subsequently, to $\rm{^4He}$, which results in the gradual relaxation of $T_R$, reaching up $10^{10}$ GeV for masses around $100$TeV.

Such limits on the reheating temperature provide important insights into the primordial universe, as they determine the maximum scale for the transition between the inflationary phase and the radiation-dominated era, impacting the generation of thermal relics such as WIMPs and other dark matter candidates \cite{Malpartida2023,Roszkowski2014}. %Therefore, the reheating temperature influences the rate of interaction of the particles present in the primordial plasma \cite{Erickcek2015}, playing an important role in determining whether dark matter candidates achieve thermal equilibrium and freeze-out at reliable abundances. Consequently, the upper bounds on the reheating temperature discussed for the gravitino decay in this section are essential for testing and refining dark matter models in the presence of supersymmetric particles. Moreover, such constraints could be used as additional information to study baryogenesis scenarios. In leptogenesis, for example, scenarios with a reheating temperature exceeding $10^9 \text{GeV}$ \cite{DAVIDSON2008} might be achievable in the presence of very heavy, unstable gravitinos.

\section{\label{Conclusions} Discussion and conclusions }

We have revisited the BBN bounds on hadronic injection episodes that took place between $\tau= 10^{-2}-10^{12}$~s based on the primordial abundances of Helium-3, Helium-4, Deuterium, and Lithium-7. Helium-4 and Deuterium provide the strongest constraints. Later we applied our limits to the gravitino problem. As BBN is sensitive to the amount of energy injected in to the plasma at a given time, BBN can constrain the primordial abundance. Knowing that the gravitino abundance is set by the reheating temperature, which is the temperature that sets the transition from inflationary phase to radiation domination, we can use BBN as probe for the reheating temperature which is of paramount importance to production mechanics of dark mater relics. The precise bound on the reheating temperature depends on the decay channel. If the gravitino decays into gluon-gluino pairs, we find $T_R< 5 \times 10^5$~GeV. This bound is significantly relaxed if the photon-photino final state is dominant.

%the reheating temperature is constrained to $ 5\times10^5 \mathrm{GeV}\lesssim T_R \lesssim \times10^7 \mathrm{GeV}$ for gravitino masses below $7$TeV. Up to the weak scale, ${}^3$He/D imposes the dominant constraint on $T_R$. After that, the reheating temperature is limited by D/H observations. For masses greater than $7$TeV, on the other hand, the constraints on the reheating temperature can be gradually relaxed. The results are similar when the gluon-gluino decay channel is kinematically blocked and the primary contribution to the gravitino decay comes from the photon-photino pair ($B_h=10^{-3}$). The key difference is that the upper limits on $T_R$ derived from ${}^3$He/D now extend to gravitino masses well beyond the weak scale, reaching up $m_{3/2}\lesssim$ 2TeV. Moreover, the constraints from D/H and $Y_p$ are much less stringent than those coming from the gluon-gluino decay. However, the presence of colored particles heavier than the gravitino mass is highly unlikely at this scale.

\acknowledgments

The authors thank Jacinto Paulo Neto and Diego Cogollo discussions. The authors acknowledge the use of the IIP cluster ''bulletcluster''. 
FSQ is supported by Simons Foundation (Award Number:1023171-RC), FAPESP Grant 2018/25225-9, 2021/01089-1, 2023/01197-4, ICTP-SAIFR FAPESP Grants 2021/14335-0, CNPq Grants 307130/2021-5 and 403521/2024-6, and ANID-Millennium Science Initiative Program ICN2019\textunderscore044, and FINEP under the project 213/2024. LA acknowledges the support from Coordenação de Aperfeiçoamento de Pessoal de Nível Superior (CAPES) under grant 88887.827404/2023-00. MMAP benefits from the support of the Conselho Nacional de Desenvolvimento Científico e Tecnológico (CNPq) under grant number 151811/2024-5.

\appendix
\section{}
\begin{figure}[htbp!]
    \centering
    \includegraphics[width=1\linewidth]{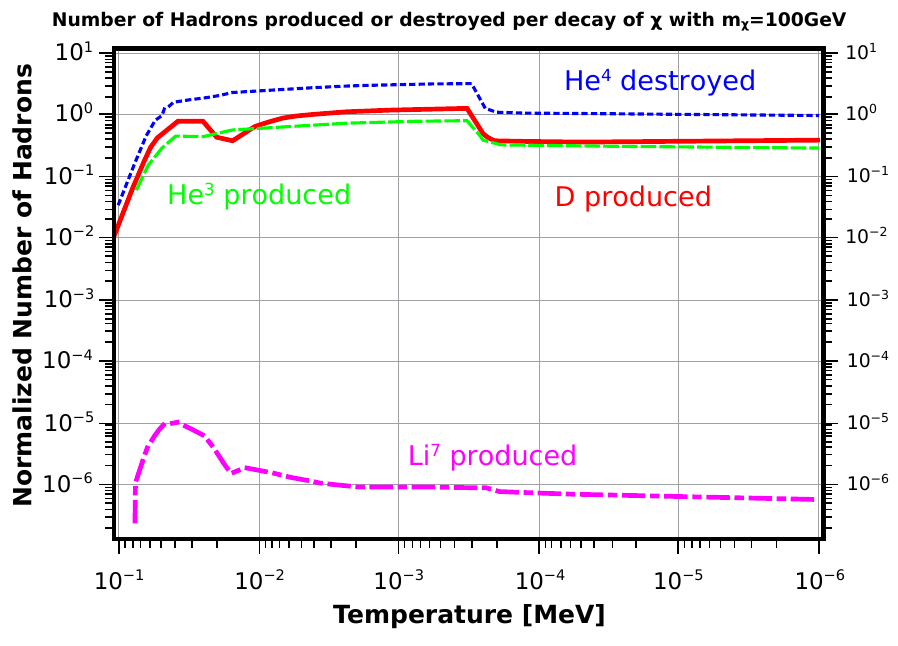}
    \caption{Number of hadrons produced or destroyed per decay of $\chi$ using $m_\chi=100$ GeV }
    \label{fig:referee}
\end{figure}

In \autoref{fig:referee} we show the number of destroyed $\rm{^4He}$ and produced $D$, $\rm{^3He}$ and $\rm{^7Li}$ as a function of the temperature. The temperatures $T=10^{-1}, 10^{-2},10^{-3}$~MeV correspond to $\tau=10^2, 10^4,10^6$~s, respectively. We fixed $m_{3/2}=100$~GeV. It is clear that for $\tau> 10^2$ s, the destruction of $\rm{^4He}$ plays a major role. The larger the mass of the decaying particle the larger the number of hadrons produced or destroyed.

\def\bibsection{\section*{References}}

\bibliographystyle{JHEPfixed.bst}
\bibliography{references}

\end{document}